\documentclass[10pt,onecolumn]{iopart}

\usepackage{iopams}  
\usepackage{graphicx}
\usepackage{bm}
\usepackage{xcolor}
\usepackage{braket}
\usepackage{url}
\usepackage{hyperref}
\begin{document}

\title[Wigner molecules and hybrid qubits]{Wigner molecules and hybrid qubits}

\author{Constantine Yannouleas \& Uzi Landman}

\address{School of Physics, Georgia Institute of Technology,
             Atlanta, Georgia 30332-0430}
\ead{Constantine.Yannouleas@physics.gatech.edu}
\ead{Uzi.Landman@physics.gatech.edu}
\vspace{10pt}
\begin{indented}
\item[]December 2021
\end{indented}

\begin{abstract}
It is demonstrated that exact diagonalization of the microscopic many-body Hamiltonian via 
systematic full configuration-interaction (FCI) calculations is able to predict the spectra as 
a function of detuning of three-electron hybrid qubits based on GaAs asymmetric double quantum 
dots. It is further shown that, as a result of strong inter-electron correlations, these 
spectroscopic patterns, including avoided crossings between states associated with different 
electron occupancies of the left and right wells, are inextricably related to the formation of 
Wigner molecules. These physical entities cannot be captured by the previously employed 
independent-particle or Hubbard-type theoretical modeling of the hybrid qubit. We report 
remarkable agreement with recent experimental results. Moreover, the present FCI methodology 
for multi-well quantum dots can be straightforwardly extended to treat Si/SiGe hybrid qubits, 
where the central role of Wigner molecules was recently experimentally confirmed as well. 
\end{abstract}

\vspace{2pc}
~~~~~~\\
Letter: J. Phys.: Condens. Matter {\bf 34}, 21LT01 (2022)\\
\url{https://doi.org/10.1088/1361-648X/ac5c28}
\maketitle


\section{Introduction}

Effective design and optimal control of the operational manipulations and interplay between the various 
degrees of freedom defining single qubit gates, as well as multi-qubit architectures, are imperatives 
for efforts targeting the successful fabrication and implementation of quantum computing devices. To 
this aim major world-wide experimental endeavors 
(see, e.g., Refs.\ \cite{kouw07,dzur13,marc10,dzur10,yaco18,marc13})
have been undertaken during the last decade. This resulted in unprecedented progress in the development 
and employment of techniques for control and manipulation of the spin and charge which serve to 
characterize two-dimensional (2D) semiconductor-based three-electron hybrid-double-quantum-dot (HDQD) 
qubits \cite{copp12,copp14.2,copp15,cao16,copp17,cao17}. Nonetheless, several recent experimental 
scrutinies on Si/SiGe \cite{erca21} and GaAs \cite{kim21} HDQD qubit devices provided unambiguous 
evidence (see also Refs.\ \cite{erca21.2,urie21}) for the need to account, in modeling the qubit physics 
and performance, for the heretofore overlooked, but unavoidable, formation of Wigner molecules (WMs)
\cite{yann99,grab99,yann00,yann00.2,boni01,yann02,mikh02,yann06,yann07,ront08}, resulting from strong 
inter-electron (e-e) interactions, and the consequent rearrangement of the spectra of the qubit device 
with respect to that associated with non-interacting electrons.  

The formation of WMs is outside the scope of investigations anchored in the framework 
of independent-particle (single-particle) modeling \cite{copp12,copp14,vinc15,cao17}, invoked at the very
early stage of studies on 2D quantum dots (QDs) \cite{marc96}. Nor are more involved Hubbard-type models 
\cite{copp12,staf94,loss99,dass11,dass11.2,ferr14,burk17} adequate for the description of the 
formation of WMs and their physical consequences. Instead, it has been demonstrated in earlier 
theoretical treatments \cite{yann99,yann00,yann00.2,yann02,mikh02,yann06,yann07,yann07.2,yann09}  
that the formation of WMs requires the employment of more comprehensive approaches, such as the
symmetry-breaking/symmetry-restoration \cite{yann99,yann00,yann02,yann07} approach or
the full configuration-interaction (FCI) method (referred to also as exact diagonalization
\cite{yann00.2,yann06,yann07,yann07.2,yann09})\footnote{%%%
For a detailed discussion of these two methodologies in the context of QDs, see the review article in 
Ref.\ \cite{yann07}.}. 

Here, motivated by the recent advances \cite{kim21,erca21,copp14.2,copp15,cao16,copp17,cao17} in the 
fabrication of charge-spin HDQD qubits, we investigate the many-body spectra and wave functions of 
three electrons in an asymmetric two-dimensional double-well external confinement, implemented by 
a two-center-oscillator (TCO) potential \cite{yann99,yann02,yann09}. In particular, we demonstrate the 
defining role that WM formation (associated with strong e-e correlations) play in shaping the 
spectra (including the key feature of a pair of left-right electron-occupancy-dependent avoided 
crossings) of semiconductor qubits by presenting the first FCI calculations for the case of a hybrid 
\cite{erca21,kim21,copp12,copp14,copp14.2,copp15,cao16,copp17,cao17} three-electron double-dot 
GaAs qubit with parameters comparable to those in Ref.\ \cite{kim21}.   

Earlier fabricated GaAs QDs \cite{marc96,vinc15,kouw01} were characterized by harmonic confinements 
with frequencies $\hbar \omega_0 \geq 3$ meV [with $R_W < 1.97$; see Eq.\ (\ref{rw})], which correspond
to a range of smaller QD sizes that did not favor the observation of the WMs at zero magnetic 
fields \cite{kouw01}. The much larger anisotropic GaAs double dot of Ref.\ \cite{kim21},
as well as the findings of Ref.\ \cite{erca21} concerning Si/SiGe dots, where strong WM signatures
were observed, herald the exploration of heretofore untapped potentialities in the fabrication and 
control of QD qubits, an objective that the present paper aims to facilitate from a theory perspective.

\section{Results}

{\it Many-body Hamiltonian:\/}
We consider a many-body Hamiltonian for $N$ confined electrons of the form
\begin{equation}
{\cal H}_{\rm MB} ({\bf r}_i,{\bf r}_j) =\sum_{i=1}^{N} H_{\rm TCO}(i) +
\sum_{i=1}^{N} \sum_{j>i}^{N} \frac{e^2}{\kappa |{\bf r}_i-{\bf r}_j|},
\label{mbhd}
\end{equation}
where ${\bf r}_i$, ${\bf r}_j$ denote the vector positions of the $i$ and $j$ electron,
and $\kappa$ is the dielectric constant of the semiconductor material. 

The single-particle $H_{\rm TCO}$ \cite{yann99,yann02,yann09}, with the unindexed coordinates
x and y corresponding to the confined particles [$i=1,\ldots,N$ in Eq.\ (\ref{mbhd})], is given by:

\begin{equation}
%\begin{eqnarray}
H_{\rm TCO}=\frac{{\bf p}^2}{2 m^*} + \frac{1}{2} m^* \omega^2_y y^2
+ \frac{1}{2} m^* \omega^2_{x k} x^{\prime 2}_k + V_{\rm neck}(x^\prime_k ) +h_k,
\label{hsp}
%\end{eqnarray}
\end{equation}
where $x_k^\prime=x-x_k$ with $k=1$ for $x<0$ (left well) and $k=2$ for $x>0$ (right well), 
and the $h_k$'s control the relative depth of the two wells, with the detuning
defined as $\varepsilon=h_1-h_2$. $y$ denotes the coordinate perpendicular to the
interdot axis ($x$). The most general shapes described by $H_{\rm TCO}$ are
two semiellipses connected by a smooth neck [$V_{\rm neck}(x^\prime_k )$]. $x_1 <
0$ and $x_2 > 0$ are the centers of these semiellipses, $d=x_2-x_1$
is the interdot distance, and $m^*$ is the effective electron mass.

For the smooth neck, we use 
\begin{equation}
V_{\rm neck}(x^\prime_k ) = \frac{1}{2} m^* \omega^2_{x k} 
\Big[ {\cal C}_k x^{\prime 3}_k + {\cal D}_k x^{\prime 4}_k \Big] \theta(|x|-|x_k|),
\label{vneck}
\end{equation}
where $\theta(u)=0$ for $u>0$ and $\theta(u)=1$ for $u<0$.
The four constants ${\cal C}_k$ and ${\cal D}_k$ can be expressed via two
parameters, as follows: 
${\cal C}_k= (2-4\epsilon_k^b)/x_k$,
and
${\cal D}_k=(1-3\epsilon_k^b)/x_k^2$, 
where the barrier-control parameters 
$\epsilon_k^b=(V_{b}-h_k)/V_{0k}$
are related to the height of the targeted interdot barrier $V_{b}$ (measured from the zero point 
of the energy scale), and $V_{0k}=m \omega_{x k}^2 x_k^2/2$. We note that measured from the 
bottom of the left ($k=1$) or right ($k=2$) well the interdot barrier is $V_{b}-h_k$.

\begin{figure*}
\centering\includegraphics[width=12cm]{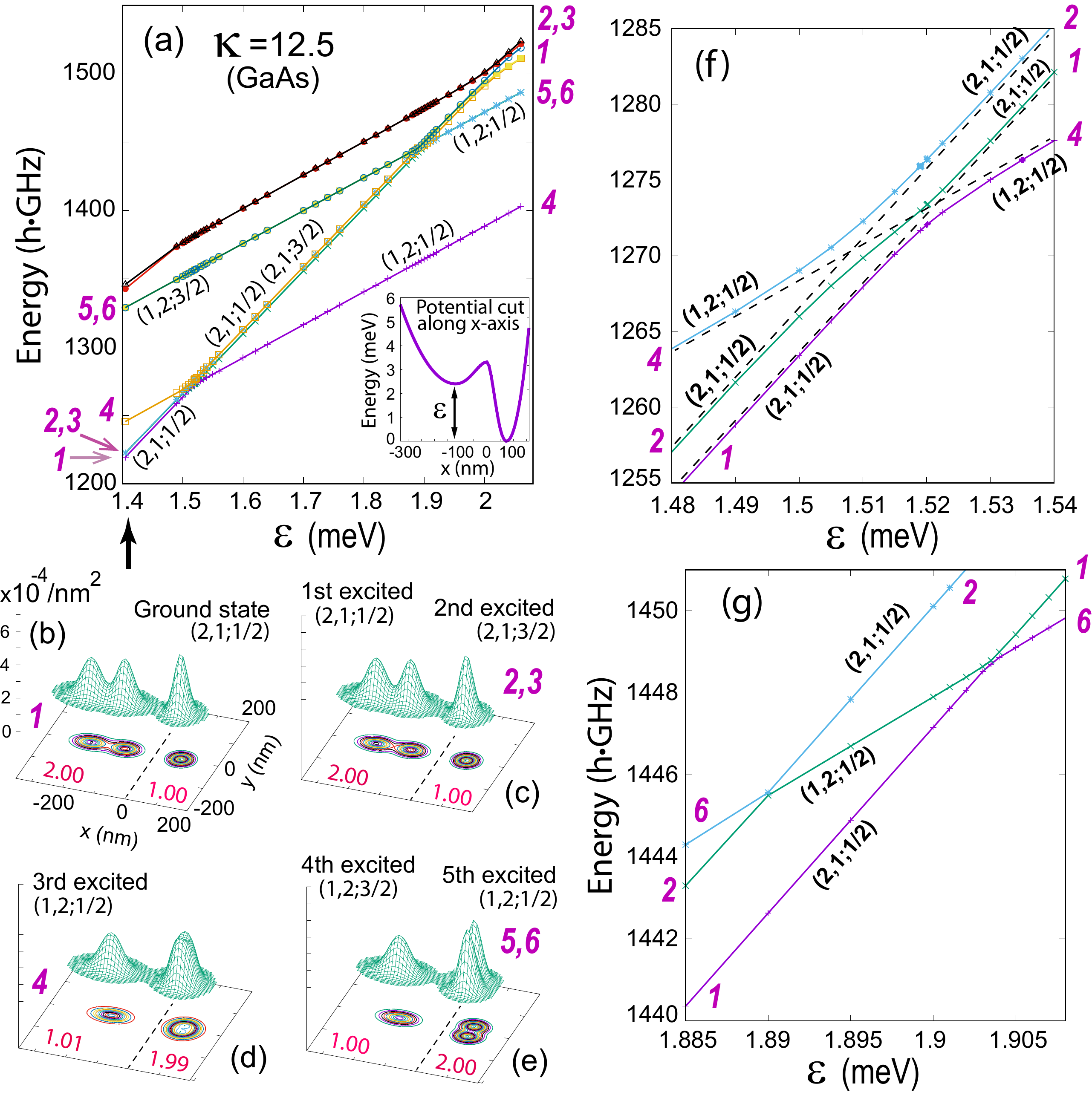}
\caption{%%%%
(a) Low-energy spectrum for the three-electron GaAs ($\kappa=12.5$) double dot.
The arrow indicates the value of the detuning at which the total charge densities were calculated. 
(b-e) Total charge densities for the ground and first five excited states. 
(f,g) Magnification of the neighborhoods of the CI avoided crossings appearing in (a).
Only the $S=1/2$ states, relevant to the hybrid qubit, are shown.
The notation $(n_L, n_R; S)$ denotes the left electron occupation, the right electron
occupation, and the total spin, respectively. For all densities, the scales of all three axes are
as in (b). CI-calculated left and right occupations are highlighted in red.
}
\label{spccp}
\end{figure*}

$H_{\rm TCO}$ has the advantage of incorporating a smooth interdot barrier $V_b$, which can be varied 
independently of the interdot separation $d$; for an illustration see the inset of Fig.\ \ref{spccp}(a).
Motivated by the asymmetric double-dot used in the GaAs device described in Ref.\ 
\cite{kim21}, we choose the parameters entering in the TCO Hamiltonian as follows: The left dot
is elliptic with frequencies corresponding to $\hbar\omega_{x1}=0.413567~{\rm meV}=100$ 
h$\cdot$GHz (long $x$-axis) and $\hbar\omega_{y1}= \hbar \omega_y = 1.22~{\rm meV}=294.9945$ h$\cdot$GHz 
(short $y$-axis), whereas the right dot is circular with $\hbar\omega_{x2}=\hbar\omega_{y2}=\hbar \omega_y =
1.22~{\rm meV}=294.9945$ h$\cdot$GHz (1 h$\cdot$ GHz $=4.13567$ $\mu$eV).
The left dot is located at $x_1=-120$ nm, and the right dot is located at $x_2=75$ nm.
The detuning parameter is defined as $\varepsilon=h_1-h_2$, where $h_1$ and $h_2$ are the chemical
potentials of the left and right dot, respectively. The interdot barrier from the bottom of the right 
dot is set to $V_b-h_2=3.3123$ meV $=800.91$ h$\cdot$GHz. Finally, the effective electron mass and the 
dielectric constant for GaAs are $m^*=0.067 m_e$ and $\kappa=12.5$, respectively.

{\it The Wigner parameter:\/}
At zero magnetic field and in the case of a single circular harmonic QD, the degree of electron
localization and Wigner-molecule pattern formation can be associated with the socalled Wigner 
parameter \cite{yann99,yann07}  
\begin{equation}
R_W =  Q/(\hbar \omega_0),                                                                           \label{rw}
\end{equation}
where $Q$ is the Coulomb interaction strength and $\hbar \omega_0$ is the energy quantum of the 
harmonic potential confinement (being proportional to the one-particle kinetic energy); 
$Q=e^2/(\kappa l_0)$, with $l_0=(\hbar/(m^*\omega_0))^{1/2}$ the spatial extension of the lowest
state’s wave function in the harmonic (parabolic) confinement.

Naturally, strong experimental signatures for the formation of Wigner molecules are not expected 
for values $R_W < 1$. In the double dot under consideration here, there are two different 
energy scales, $\hbar\omega_1=0.413567~{\rm meV}$ (associated with the long $x$ dimension of the 
left QD) and $\hbar\omega_2=1.22~{\rm meV}$ (associated with the right circular QD). As a result, 
for GaAs (with $\kappa=12.5$) one gets two different values for the Wigner parameter, namely 
$R_{W,1}=5.31$ and $R_{W,2}=3.09$. These values suggest that a stronger Wigner molecule should 
form in the left QD compared to the right QD, as indeed was found by the FCI calculation
(see below). 

%------------------------------ begin figure 1 ------------------------
%------------------------------ end figure 1 ------------------------

{\it CI spectra as a function of detuning:\/} We use the three-part notation $(n_L, n_R; S)$ to 
denote the left-well electron occupation, the right-well electron occupation, and the total spin, 
respectively; $S=1/2$ or $S=3/2$ for three electrons.

In Fig.\ \ref{spccp}(a)], the low-energy spectrum in the GaAs case $(\kappa=12.5)$ is displayed
in the range of detunings $1.40 \mbox{~meV} \leq \varepsilon \leq 2.1 \mbox{~meV}$.
The $(2,1;S)$ states with two electrons in the {\it left} well, along the $(1,2;S)$ states with 
two electrons in the right well, are prominent. States with three electrons in a given well, 
associated with a notation $(3,0;S)$ or $(0,3;S)$, are absent. The fact that only the six $(2,1;S)$ 
and $(1,2;S)$ states comprise the lowest-energy spectrum for the GaAs double dot is an essential 
feature that is a prerequisite for the implementation of the hybrid qubit which uses
\cite{copp12,copp14,copp14.2,kim21,cao17} the four $(2,1;1/2)$ and $(2,1;1/2)$ states.
As discussed below, this feature is brought about by the formation of WMs resulting from
strengthening of the typical Coulomb interaction energies relative to the energy gaps in the
single-particle spectrum of a confining external potential that represents a rather large-size
and strongly asymmetric double dot (see the earlier discussion on the Wigner parameter $R_W$).

In Fig.\ \ref{spccp}(a), we have successively numbered the lowest six states at $\varepsilon=1.4$ 
meV, starting from the ground state (\#1) and moving upwards to the first five excited ones. Apart 
from the immediate neighborhood of an avoided crossing, these energy curves are straight
lines, and naturally we extend the same numbering for all values of the detuning in the window range
used in Figs.\ \ref{spccp}(a).

The spectrum in Fig.\ \ref{spccp}(a) requires additional commentary, because of quasi-degeneracies 
between the states \#2,\#3, and \#5,\#6, as well as the small energy gap ($\sim$ 3 h$\cdot$GHz)
between state \#1 and the quasi-degenerate pair (\#2,\#3). We stress that the states \#1 and \#2 
have two electrons in the {\it left} well and total spin $S=1/2$, and thus they are denoted as
$(2,1;1/2)$, whereas state \#3 has two electrons in the left well, but a total spin of $S=3/2$
[denoted as $(2,1;3/2)$]. On the other hand, states \#4, \#6 (with $S=1/2$), and \#5 (with
$S=3/2$) have two electrons in the {\it right} well and they are denoted as $(1,2;S)$. A main feature of 
this six-state spectrum in  Fig.\ \ref{spccp}(a) is that, apart from the neighborhoods of the two 
avoided crossings, the  energy curves for the states \#1, \#2, and \#3 form one
band of parallel lines, whereas the energy curves for the states \#4, \#5, and \#6 form a second
band of parallel lines, and the two bands intersect at two avoided crossings. 

We reiterate that the appearance of such three-member bands, grouping together two $S=1/2$ states and 
one $S=3/2$ state, is a consequence of the formation of a 3e WM (three localized
electrons considering both wells), and this organization is in consonance with the findings of Ref.\
\cite{yann07.2} regarding the spectrum of three electrons in single anisotropic quantum dots in
variable magnetic fields. We further stress that the dominant feature in the spectrum shown in Fig.\ 
\ref{spccp}(a) is the small energy gap between the two $S=1/2$ states \#1 and \#2, which contrasts 
with the large gap between the other two $S=1/2$ states \#4 and \#5, a behavior that agrees with
the experimental findings of Ref.\ \cite{kim21}.

{\it Charge densities away from the avoided crossings:\/} 
Further insights into the unique trends and properties of the GaAs HDQD qubit are  gained through an 
inspection of the CI charge densities, plotted in Figs.\ \ref{spccp}(b-e) for the ground and first five 
excited states. The red numbers indicate the left-well and right-well electron occupations as calculated
from the CI method. Naturally, the charge densities are normalized to the total number of electrons 
$N=3$.

The charge densities deviate strongly from those expected from an independent-particle system. 
Indeed the formation of a strong 2e WM in the left well and of a weaker 2e WM in the right well is 
clearly seen through the emergence of a double hump in all six cases.

{\it The avoided crossings:\/} 
The position and the asymmetric anatomy of the two avoided crossings [Figs.\ \ref{spccp}(a,f,g)] play 
an essential role in the operation of the hybrid qubit \cite{kim21}, requiring a FCI simulation 
that incorporates both dots of the HDQD qubit, as demonstrated here\footnote{%%%%
The qubit is initialized in the ground-state on line \#4 (tuned to the far right of the left crossing)
in Fig.\ \ref{spccp}(a). After detuning and laser-pulse-induced jumping to state \#2 [at left crossing,
Fig.\ \ref{spccp}(f)], readout is achieved via increased detuning, moving along state \#1 and through
the right avoided crossing to state \#6 [Fig.\ \ref{spccp}(g)].}.
In Fig.\ \ref{spccp}(f) and Fig.\ \ref{spccp}(g), we display magnifications of the neighborhoods of 
the left and right CI avoided crossings, respectively, appearing 
in the spectrum of the GaAs double dot [Fig.\ \ref{spccp}(a)]. Only the $S=1/2$ states are shown, 
because the $S=3/2$ states are not relevant for the workings of the hybrid qubit 
\cite{kim21,copp12,vinc00,burk17}. 

The left avoided crossing (situated in the neighborhood of 1.49 meV $<~\varepsilon~<$ 1.54 meV)
is formed through the interaction of the three curves \#1, \#2, and \#4 [we keep the same numbering
of the curves here as in Fig.\ \ref{spccp}(a)]. On the other hand, the curves \#1, \#2, and \#6
participate in the formation of the right avoided crossing in the neighborhood of 1.885 meV
$<~\varepsilon~<$ 1.908 meV. We note that, according to the FCI calculation, the two avoided crossings
are separated by a detuning distance of $\sim 400$ $\mu$eV, which agrees with the experimentally
determined value for the hybrid qubit device in Ref.\ \cite{kim21}.

The continuous lines in Figs.\ \ref{spccp}(f) and \ref{spccp}(g) represent the socalled {\it adiabatic\/}
paths, which the system follows for slow time variations of the detuning. For fast time variations of
the detuning, or with an applied laser pulse, the system can instead follow the {\it diabatic\/} paths
indicated explicitly with dashed lines in Fig.\ \ref{spccp}(f) and thus jump from one adiabatic line to
another; this occurs according to the celebrated Landau-Zener-St\"{u}ckelberg-Majorana
\cite{cao13,copp12.2,burk13} dynamical interference theory.

%------------------------------ begin figure 2 ------------------------
\begin{figure}[t]
\centering\includegraphics[width=7.0cm]{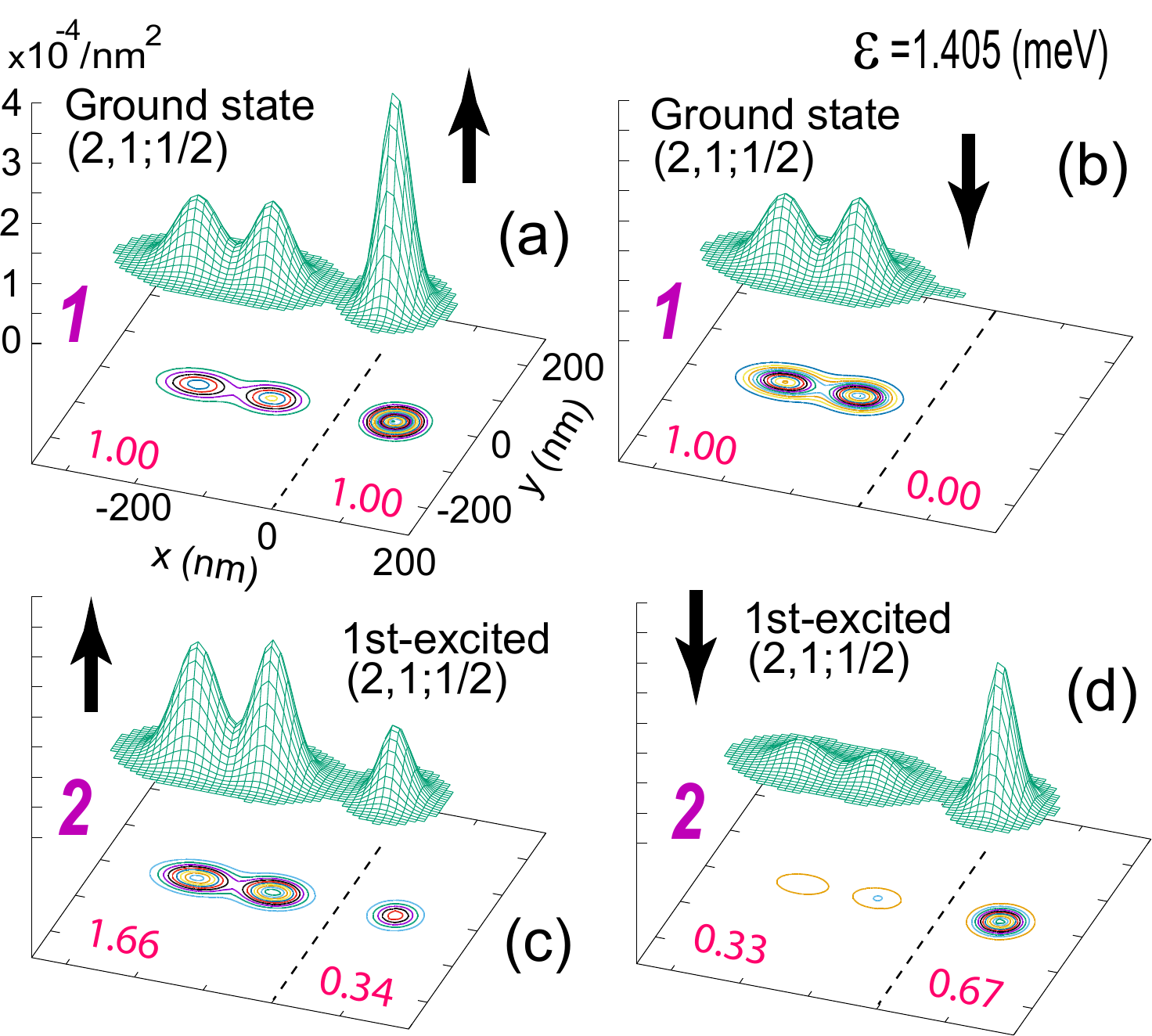}
\caption{%%%%
The spin structure of the ground state (a,b) and 1st-excited (c,d) states at $\varepsilon=1.405$ 
[see Figs.\ \ref{spccp}(b) and \ref{spccp}(c) for the corresponding total charge densities]. 
The red numbers indicate the CI-calculated left and right occupancies (rounded to the 
second decimal point). The spin-resolved densities integrate to the number of spin-up and spin-down 
electrons in (a) and (b), respectively. The arrows indicate the spin direction.
}
\label{spnstr1}
\end{figure}
%------------------------------ end figure 2 ------------------------

{\it Spin structure away from the avoided crossings:} 
The charge densities associated with the states 
\#1, \#2, and \#3 in the three-member band are designated with the same numbers and are plotted in the 
top two frames of Fig.\ \ref{spccp}(b). These three charge densities are very similar.
However the corresponding spin structures are different. We analyze below the two cases of the 
ground state and the 1st-excited state for $\varepsilon=1.405$ meV.

Fig.\ \ref{spnstr1}(a) and Fig.\ \ref{spnstr1}(b) display the spin-up and spin-down densities for
the ground state mentioned above; compare Fig.\ \ref{spccp}(b) for the total charge density. From
these two spin-resolved densities, it is immediately seen that the spin structure of this ground 
state conforms to the following familiar expression \cite{vinc00,burk17,copp12,kim21,yann07.2}
in the theory of three-electron qubits and quantum dots:
\begin{equation}
(|duu\rangle-|udu\rangle)/\sqrt{2},
\label{spin1}
\end{equation}
where $u$ and $d$ denote an up and down spin, respectively, with the three spins arranged from left 
to right in three ordered sites.

Fig.\ \ref{spnstr1}(c) and Fig.\ \ref{spnstr1}(d) display the spin-up and spin-down densities for
the associated 1st-excited state; compare Fig.\ \ref{spccp}(c) for the total charge density. From
these two spin-resolved densities, one can conclude that the spin structure of this 1st-excited
state conforms to a second familiar expression \cite{vinc00,burk17,copp12,kim21,yann07.2} 
in the theory of three-electron qubits and quantum dots, namely
\begin{equation}
(2|uud\rangle-|duu\rangle-|udu\rangle)/\sqrt{6}.
\label{spin2}
\end{equation}

Indeed, from Eq.\ (\ref{spin2}), one can derive that the expected spin-up occupancy for the most 
leftward and middle positions of the three spins is $5/6$ in both cases, 
yielding $5/3=1.666$ for the expected spin-up occupancy in the left dot, 
in agreement with the CI value of 1.66 highlighted in red in Fig.\ \ref{spnstr1}(c).
Similarly the expected spin-up occupancy for the right dot from Eq.\ (\ref{spin2}) is $1/3=0.333$,
in agreement with the CI-value of 0.34 highlighted in red in Fig.\ \ref{spnstr1}(c).
Moreover, from Eq.\ (\ref{spin2}), one can derive that the expected spin-down occupancy for the most
leftward and middle positions of the three spins is $1/6$ in both cases, 
yielding $1/3=0.333$ for the expected spin-down occupancy in the left dot, in agreement with the CI 
value of 0.33 highlighted in red in Fig.\ \ref{spnstr1}(d).
Finally the expected spin-down occupancy for the right dot from Eq.\ (\ref{spin2}) is $2/3=0.666$,
in agreement with the CI-value of 0.67 highlighted in red in Fig.\ \ref{spnstr1}(d).

{\it The effective matrix Hamiltonian:\/}
From the CI spectra, one can extract the phenomenological effective matrix Hamiltonian 
\cite{kim21,copp14.2} that has played a central role in the experimental dynamical control of the 
hybrid qubit. The general form of this 4$\times$4 matrix Hamiltonian is:
\begin{equation}
H_M=\left(
\begin{array}{cccc}
c_L \widetilde\varepsilon/2 & 0 & \delta_1 & -\delta_2 \\
0 & c_L \widetilde\varepsilon/2 + \Delta E_L & -\delta_3 & \delta_4 \\
\delta_1 & -\delta_3 & c_R \widetilde\varepsilon/2 & 0\\
-\delta_2 & \delta_4 & 0 & c_R \widetilde\varepsilon/2+\Delta E_R,
\end{array}
\right) 
\label{eqtoyham}
\end{equation}
where $\widetilde\varepsilon=\varepsilon-\varepsilon_0$ here denotes a renormalized detuning.

A good fit with the CI spectrum in Figs.\ \ref{spccp}(a), \ref{spccp}(f), and \ref{spccp}(g) is 
achieved by setting $c_L=4.4$, $\Delta E_L=15$ $\mu$eV, $c_R=2.7$, $\Delta E_R=340$ $\mu$eV, 
$\delta_1=0.657$ $\mu$eV, $\delta_2=0.090$ $\mu$eV, $\delta_3=1.207$ $\mu$eV, $\delta_4=0.075$ $\mu$eV,
and $\varepsilon_0=1.50$ meV. 

The effective matrix Hamiltonian in Eq.\ (\ref{eqtoyham}) reflects (within the plotted 
window) two properties of the FCI spectrum in Fig.\ \ref{spccp}(a) that are instrumental 
\cite{cao17.2,copp14.2} for the successful operation of the hybrid qubit, namely, the quasi-linear 
dependence of $H_M$ on the detuning $\widetilde\varepsilon$ and the quasi-parallel
behavior of both the two $(2,1,1/2)$ states (states \#1 and \#2) and the two $(1,2,1/2)$ states 
(states \#4 and \#6). We note a difference between Refs.\ \cite{kim21,copp14.2} and
the CI result for $H_M$, namely, Refs.\ \cite{kim21,copp14.2} assume the values $c_L=1$ and $c_R=-1$
associated with 45$^o$ and -45$^o$ slopes of the associated lines, respectively, while the CI 
result produces different slopes associated with  $c_L=4.4$ and $c_R=2.7$. 

\section{Conclusions}

We presented extensive FCI results that combine both energetics and investigation of the 
many-body wave functions through the calculation of charge and spin-resolved densities. 
Going beyond two-particle CI treatments in a single dot \cite{erca21,erca21.2,kim21,urie21}, 
this paper enabled for the first time 
the investigation of key features appearing in the low-energy spectrum of actual experimentally 
fabricated GaAs three-electron HDQD qubits, and in particular the role of a pair of avoided crossings 
between levels corresponding to different electron occupancies in the left and right wells. 
We demonstrated that the emergence of these spectral features, which are codified in a simple 
effective matrix Hamiltonian [Eq.\ (\ref{eqtoyham})], emanating from the complexity of the many-body 
problem, is inextricably related to electron localization and the formation of Wigner molecules. 

Derivations \cite{copp12,ferr14} of the matrix Hamiltonian in Eq.\ (\ref{eqtoyham}), starting from 
approximate two-site Hubbard-type modeling, involve qualitative approximations which are not applicable 
for the case of WM formation. Consequently, the present CI-based confirmation of this effective matrix 
Hamiltonian, accounting fully for strong-correlation effects within each well and WM formation, is an 
unexpected auspicious result.  

Our multi-dot FCI method can straightforwardly be expanded to incorporate the valley degree of freedom,
thus holding the potential for being adopted as an effective tool for analysing and designing hybrid 
qubits, including the case of Si/SiGe hybrid qubits with more than three electrons where more complex
spectra have been recently experimentally discovered \cite{erca21}.
In this context, a main focus of ongoing research \cite{yann22} is the investigation of the effect 
of the valley degree of freedom on the formation of near-degenerate pairs of electronic states. 
We mention again that, 
in the case of the GaAs qubit device \cite{kim21} considered here, the quasi-degeneracies is an effect 
of the strong $e-e$ interaction and Wigner-molecule formation, which suppress the energy gaps in the
electronic spectrum. The valley degree of freedom will introduce further possibilities for
grouping of the electronic states of the qubit device due to additional group symmetries that
become apparent when the valley-pseudospin analogy is explicitly considered; e.g., the $SU(4)$ or 
$SU(2) \times SU(2)$ group symmetries.

\ack
This work has been supported by a grant from the Air Force Office of Scientific Research (AFOSR) 
under Award No. FA9550-21-1-0198. Calculations were carried out at the GATECH Center for 
Computational Materials Science.

\section*{References}
\bibliography{DoubleDot_3e_asymm_GaAs_JPCM}

\end{document}